\documentclass[12pt,fleqn]{article}

\usepackage{comment}

\usepackage{graphicx} 

\usepackage{setspace}
\usepackage[margin=1in]{geometry}
\usepackage[symbol]{footmisc}

\usepackage[T1]{fontenc}
\usepackage[utf8]{inputenc}
\usepackage{authblk}

\makeatletter

\newcommand{\fboxsubsec}[1]{
	\begin{flushleft}
		#1
	\end{flushleft}
	}
\renewcommand{\subsection}{\@startsection{subsection}{2}{0pt}
	{1ex}
	{0.5ex}
	{\reset@font\it\fboxsubsec}
	}
\makeatother

\title{Detecting Structural Breaks in Foreign Exchange Markets by using the group LASSO technique\footnote{The author reported the previous version in the 96th meeting of Western Economic Association International in 2021.
The author received useful suggestions form Professor Tatsuma Wada in the meeting. Mr. Akio Sashida suggested several possible events that might trigger breaks within my sample period. }}%

\author{Mikio Ito \\{\scriptsize  \it Keio University, 2-15-45 Mita, Minato-ku, Tokyo 108-8345, Japan}}

\date{\today}

\usepackage[]{natbib}%
\usepackage{amsmath,amssymb}%
\usepackage{ascmac}%
\usepackage{multirow}%
\usepackage{lscape}%
\usepackage{subfigmat}

\usepackage{pifont}%
\usepackage{arydshln}%
\usepackage[format=hang]{caption}
\usepackage[all]{xy}
\usepackage{url}
\bibpunct{(}{)}{;}{a}{}{,}

\def\hsymbu#1{\smash{\lower1.7ex\hbox{\huge$#1$}}}

\def\ve #1{{\mbox{\boldmath $#1$}}}

\def\ve #1{{\mbox{\boldmath $#1$}}}

\begin{document}

\maketitle


\noindent

\noindent
{\bfseries Abstract:} This article proposes an estimation method to detect breakpoints for linear time series models with their parameters that jump scarcely. Its basic idea owes the group LASSO (group least absolute shrinkage and selection operator). The method practically provides estimates of such time-varying parameters of the models. An example shows that our method can detect each structural breakpoint's date and magnitude.
\\



\noindent






\setcounter{equation}{0}
\renewcommand{\theequation}{\arabic{equation}}
\renewcommand{\thetable}{\arabic{table}}
\renewcommand{\thefigure}{\arabic{figure}}
\renewcommand{\thesection}{\arabic{section}}
\renewcommand{\thesubsection}{\arabic{subsection}}

\newenvironment{proof}[1][Proof]{\textbf{#1.} }{\ \rule{0.5em}{0.5em}}
\let\tilde=\widetilde

\newpage

\section{Introduction}\label{lab:introduction}
Recent development in data science has enabled us to handle 'big' datasets, including high dimensional time series; such data have been increasingly common in economics and finance. At the same time, we can adopt models with more than ten thousand parameters to estimate, as this paper addresses. A sparse regression technique based on the least absolute shrinkage and selection operator (LASSO) or a similar one is typical to extract useful information from such big data. (See 
\citet{tibshirani2015statistical}
for general discussion on LASSO.)

There exists recent literature about the theory and application of the LASSO technique to time-series datasets.   Articles by 
\cite{michailidis2013autoregressive}
and 
\cite{basu2015regularized}
study vector autoregressive model using the LASSO technique. 
\cite{chan2014group}
adopt a group LASSO model to solve the problem in a variable selection for a structural break autoregression model
\cite{ito2019LASSO}
shows a method to estimate state parameters that jump rarely based on the group LASSO. 

On the other hand, recent economists use econometric models with parameters varying over time when they are interested in structural breaks of economic or financial systems. For example, Primiceri(2005) studies a macroeconomic model with parameters varying over time by adopting a Bayesian technique to estimate a state-space model. The author and his coworkers apply a conventional least square method to estimate time-varying parameters of time series models such as a VAR model. (See
\cite{ito2021time}.) 

This article proposes an estimation method to detect breakpoints for linear time series models with their parameters that jump scarcely.    One usually supposes non-Gaussian processes like the combination of Gaussian and Poisson ones for the models; the author supposes no statistical specific noises in his estimation. His basic idea owes a sparse regression technique to regard them as linear regression models with possibly quite many parameters. Its most significant point is that scarcely changes in state variables are equivalent to a sparse structure of solutions for a transformed linear equation system obtained from the linear time series model by differencing parameters. The transformation allows the researchers to employ group LASSO techniques when they estimate quite many parameters supposed to jump.

The article's organization is as follows. Section 2 reviews a basic idea. It shows how to transform a linear time series model into another with differenced parameters allowing group LASSO techniques. Finally, the section proposes the practical method to estimate a time series model with multiple parameters scarcely jumping over time. Section 3 shows an application of how we detect several structural changes in foreign exchange markets by using our method. In practice, the author employs the technique for the same data that he and his co-authors studied the foreign exchange markets in \cite{ito2021time}. The result shows the periods in which structural changes happened in the markets in a sharper way than \cite{ito2021time}. Section 4 concludes and shows several points for future research tasks.

\section{Linear Time series Model with parameters that jump scarcely}\label{sec:model}
There is literature about the estimation of time-varying coefficients of economic models. P. Swamy has published many articles on broadly theoretical frameworks providing several models with time-varying parameters since the 1970's, for example, \cite{swamy2010estimation}. However, we focus on a narrower class of models.
Suppose $m\times n$  matrix $X_t$ and $m$-dimensional vector $\ve{y}_t$ are  given at each $t$. 
\begin{equation}
\ve{y}_t = X_t \ve{\beta}_t + \ve{u}_t, \quad (t=1\cdots T) \label{eq:obs 0}, 
\end{equation}
where $\ve{\beta}_t$ is a $n$-dimensional vector coefficient to be estimated. Let $t$ denote time. Thus $T$ is the sample size when the data is set.

We can regard the above model as a linear regression model whose each coefficient $\ve{\beta}$ varies over time. At this time, we do not have to consider $\ve{u}_t$'s are gausian. We can rewrite (\ref{eq:obs 0}) in the following matrix form.
\begin{equation}   
\left[
\begin{array}{c}
    \ve{y}_{ 1 } \\
    \ve{y}_{ 2 } \\
    \ve{y}_{ 3 } \\
    \vdots   \\
    \ve{y}_{ T } 
\end{array}
\right]
=
\left[
\begin{array}{ccccc}
    X_{ 1 } & & & & \Large{\ve{O}}\\
    & X_{ 2 } & & &  \\
    & & X_{ 3 } & & \\
    & & & \ddots &   \\
    \Large{\ve{O}} & & & & X_{ T } 
\end{array}
\right]
\left[
  \begin{array}{c}
    \ve{\beta}_1\\
    \ve{\beta}_2\\
    \ve{\beta}_3\\
    \vdots \\
	\ve{\beta}_T  
  \end{array}
\right]
+
\left[
  \begin{array}{c}
    \ve{u}_1\\
    \ve{u}_2\\
    \ve{u}_3\\
    \vdots \\
	\ve{u}_T  
  \end{array}
\right]
\label{eq: matrix form (1)}
\end{equation}
The data matrix in the regression (\ref{eq: matrix form (1)}) is $mT \times nT$ and the coefficient vector is of $nT$ dimension. Thus, usual regression solver such as OLS is inapplicable since the coefficient matrix in (\ref{eq: matrix form (1)}) is singular.

Now,we suppose that (\ref{eq:obs 0}) is a regression model whose coefficients $\ve{\beta}_t$'s quite rarely change. Then, we rewrite the linear equation system (\ref{eq:obs 0}) into another linear system with respect to $\Delta\ve{\beta}_t$ in place of $\ve{\beta}_t$ to focus on sparsity. The following relations are my key idea.
\begin{equation}
	\left\{
    \begin{array}{l}
    \ve{\beta}_1 = \ve{\beta}_0 + \Delta\ve{\beta}_1 \\
    \ve{\beta}_2 = \ve{\beta}_0 + \Delta\ve{\beta}_1 + \Delta\ve{\beta}_2   \\
               \vdots     \\
    \ve{\beta}_T = \ve{\beta}_0 + \Delta\ve{\beta}_1 + \Delta\ve{\beta}_2 + \cdots + \Delta\ve{\beta}_T,   \\
    \end{array}
	\right.
\label{eq: diffs}
\end{equation}
where we regard $\ve{\beta}_0$ as the starting value given of a sequence of $\Delta\ve{\beta}'s.$ 

From (\ref{eq: diffs}), we obtain its matrix form.
\begin{equation}\label{eq: diffs matrix form}
\left[
\begin{array}{c}
    {\ve{\beta}}_{ 1 } \\
    {\ve{\beta}}_{ 2 } \\
    {\ve{\beta}}_{ 3 } \\
    \vdots   \\
    {\ve{\beta}}_{ T } 
\end{array}
\right]
=
\left[
  \begin{array}{ccccc}
    I & \ve{O} & \ve{O} & \cdots & \ve{O} \\
    I & I &  \ve{O} & \cdots & \ve{O} \\
    I & I & I & {} & \Large{\ve{O}} \\
    \vdots & \vdots & \vdots & \ddots & {} \\
    I & I & I & \cdots & I
  \end{array}
\right]
\left[
\begin{array}{c}
    \Delta{\ve{\beta}}_{ 1 } \\
    \Delta{\ve{\beta}}_{ 2 } \\
    \Delta{\ve{\beta}}_{ 3 } \\
    \vdots   \\
    \Delta{\ve{\beta}}_{ T } 
\end{array}
\right]
+
\left[
\begin{array}{c}
    {\ve{\beta}}_{ 0 } \\
    {\ve{\beta}}_{ 0 } \\
    {\ve{\beta}}_{ 0 } \\
    \vdots   \\
    {\ve{\beta}}_{ 0 } 
\end{array}
\right]
\end{equation}
One can set $\ve{\beta}_0$ as the coefficient vector for the usual time-invariant regression. (Please see Appendix for detail.)

We apply the group LASSO method to the following equation.
\begin{equation}\label{eq:sprase TVP regression}
\ve{r} = \ve{\tilde X} \Delta\ve{\beta} + \ve{u},
\end{equation}
where $\ve{r}$ is the residuals of a VEC model with time invariant parameters and $\ve{\tilde X}$ is the matrix as shown in the following two slides.

\begin{equation}   
\ve{r}
=
\left[
\begin{array}{c}
    {\ve{r}}_{ 1 } \\
    {\ve{r}}_{ 2 } \\
    {\ve{r}}_{ 3 } \\
    \vdots   \\
    {\ve{r}}_{ T } 
\end{array}
\right]
=
\left[
\begin{array}{c}
    \ve{y}_{ 1 } \\
    \ve{y}_{ 2 } \\
    \ve{y}_{ 3 } \\
    \vdots   \\
    \ve{y}_{ T } 
\end{array}
\right]
-
\left[
\begin{array}{ccccc}
    X_{ 1 } & & & & \Large{\ve{O}}\\
    & X_{ 2 } & & &  \\
    & & X_{ 3 } & & \\
    & & & \ddots &   \\
    \Large{\ve{O}} & & & & X_{ T } 
\end{array}
\right]
\left[
  \begin{array}{c}
    \ve{\beta}_0\\
    \ve{\beta}_0\\
    \ve{\beta}_0\\
    \vdots \\
	\ve{\beta}_0  
  \end{array}
\right]
\label{eq:bar y}
\end{equation}
and
\begin{equation}   
\tilde{\ve{X}}
=
\left[
\begin{array}{ccccc}
    X_{ 1 } & & & & \Large{\ve{O}}\\
    & X_{ 2 } & & &  \\
    & & X_{ 3 } & & \\
    & & & \ddots &   \\
    \Large{\ve{O}} & & & & X_{ T } 
\end{array}
\right]
\left[
  \begin{array}{ccccc}
    I & \ve{O} & \ve{O} & \cdots & \ve{O} \\
    I & I &  \ve{O} & \cdots & \ve{O} \\
    I & I & I & {} & \Large{\ve{O}} \\
    \vdots & \vdots & \vdots & \ddots & {} \\
    I & I & I & \cdots & I
  \end{array}
\right],
\label{eq:tilde X}
\end{equation}
Where $I$ is the $n\times n$ identity matrix.

At this point, we provide our breakpoint detecting procedure. 
\begin{description}
\item[Step 1] We apply a group Lasso solver to the equation (\ref{eq:sprase TVP regression}) by grouping variables according to periods.
  \begin{itemize}
    \item { Both ${\ve{r}}$ and each column of $\tilde{\ve{X}}$ should be "normalized" before the above operation.}
  \end{itemize}
\item[Step 2] The group Lasso solver finds a series of $\Delta\ve{\beta}_t$ whose all components are zeros for almost all periods. 
\item[Step 3] We regard the period such that $t$ such that $\Delta\ve{\beta}_t \ne \ve{0}$ as a break point.
\end{description}
Note that we can obtain each $\ve{\beta}_t$ through the following equation 
$$\ve{\beta}_t=\hat{\ve{\beta}}_0 + \sum_{\tau=1}^{t}\Delta\ve{\beta}_\tau.$$

We call the linear equation system (\ref{eq:sprase TVP regression}) a {\bf difference form} of the linear time series model. Note that the transformation above from the original model (\ref{eq:obs 0}) to the different form is quite algebraic and that the error terms are identical. Thus, the linear equation (\ref{eq:sprase TVP regression}) as a statistical model is equivalent to the original state-space model corresponding to the equation (\ref{eq:obs 0}). Irrespective of an estimation method, the above two statistical models provide essentially the same results. One should contemplate which kind of noises they choose, Gaussian or non-Gaussian, when they employ a linear time series model given data $\ve{y}$ observed.  

As the author stressed in \cite{ito2019LASSO}, when we address state-space models covering the linear time series model,  we have assumed Gaussian noises in many applications. At the same time, their error terms are not always supposed to follow Gaussian noises. Intuitively, under the assumption of Gaussian noise for the models, the sequence of estimated parameters tends to be smooth. In contrast to Gaussian noises, under the assumption of non-Gaussian noise, for instance, the Poisson process for a state-space model, the sequence of estimated state variables tends to be almost invariant and jump scarcely. 

Potentially supposing broader classes of errors,  the author regards LASSO (or other sparse regression solvers) as a suitable solver for a linear time series model with parameters that jump in scarce timings. LASSO effectively solves the corresponding difference form (\ref{eq:sprase TVP regression}) even if the number of parameters amounts to more than ten thousand, as is shown in the next section. In this case, the group LASSO or generalized LASSO is applicable. (See \cite{yuan2006model} and \cite{tibshirani2011solution} for detail of the two LASSO.)

\section{Application to Foreign Exchange Markets}       
The author shows an application in which the method in Section \ref{sec:model} possibly detects several structural changes in a real financial market. The data is the same as \cite{ito2021time}. They used
monthly nominal data from the Thomson Reuters Datastream on the spot and forward exchange rates for three developed countries (Canada, Japan, and the UK) from May 1990 to July 2015 taken. (See Ito et al.(2021) for detail) They estimated a time-varying VEC model's parameters,  supposing them varying like random walks over time. They argue that the estimated degree of market comovement has increased over sample periods. At the same time, the behavior is not monotonous. One can regard the degree as the speed of adjustment if the deviation from the long-run equilibrium exists. Figures 1 and 2 summarize their findings.

\begin{figure}
 \centering
   \includegraphics[width=100mm]{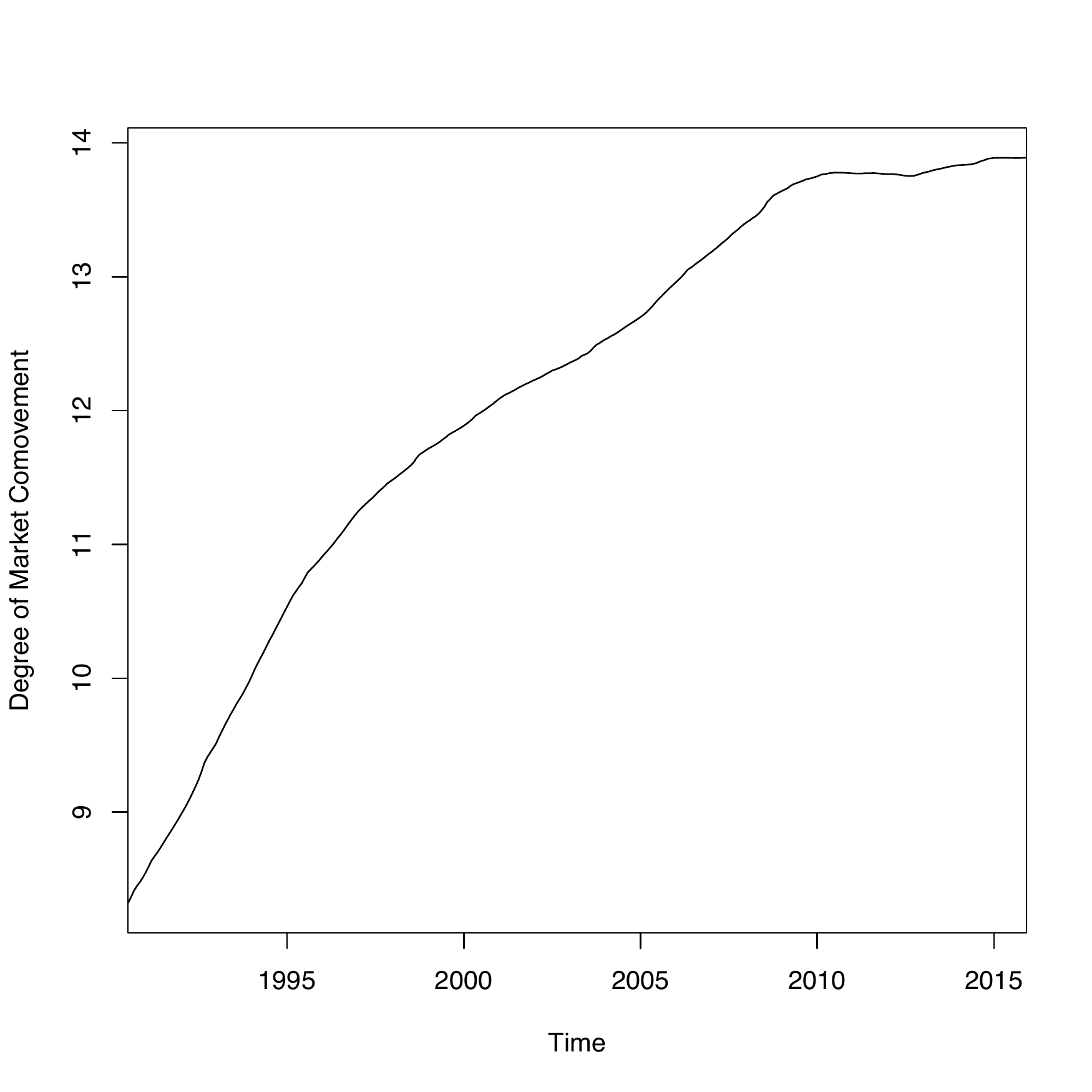}
 \caption{Degree of comovement in foreign exhange rates (1990-2015) (from \cite{ito2021time})}
 \label{fig:premium}
\end{figure}

\begin{figure}
 \centering
   \includegraphics[width=100mm]{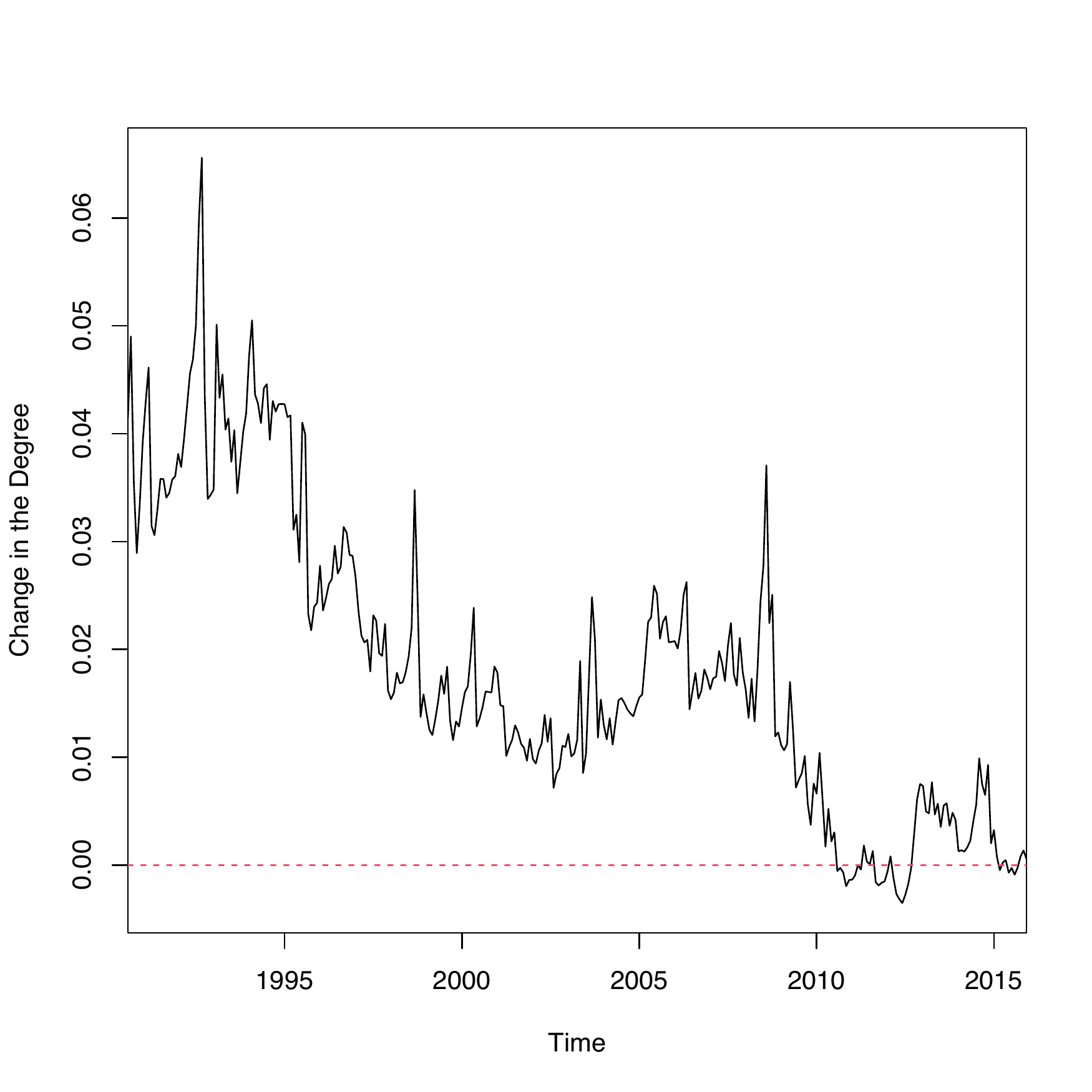}
 \caption{Changes in the Degree of comovement (1990-2015) (from \cite{ito2021time})}
 \label{fig:spot returns}
\end{figure}

\newpage

This section aims to show possible breakpoints sharper than \cite{ito2021time} and in what periods they occurred.
Ito et al.(2021) suggests that the exchange markets vary over time and that the changes are gradual.
Assuming there are some cointegrating relationships, we consider a vector error correction (VEC) model for $m$-vector time series $X_{t}$.
\begin{equation}\label{eq: VECM basics}
\Delta X_{t}=\Gamma_{1}\Delta X_{t-1}+\cdots+\Gamma_{k}\Delta X_{t-k}+\Pi_{k}X_{t-k}+\mu +\ve{u}_{t},  
\end{equation}
where $\Delta X_{t}=X_{t}-X_{t-1}$, $\mu$ is a vector of intercepts, and $\ve{u}_{t}$ is a vector of error terms. 
\vspace{1cm}
The Coefficient matrix $\Pi_k$ represents some long-run relationships among the components of $X_t$'s.
Regarding (\ref{eq: VECM basics}) as a linear time series regression (\ref{eq: matrix form (1)}) and supposing $\Pi_k$ varying over time, we apply our group LASSO technique. 

At this point, we show a practical procedure applied in this section as follows.
\begin{description}
\item[Step 1] First, we estimate a usual VEC model supposing any coefficient time-invariant after some cointegration test.
\item[Step 2] Using its loading matrix $\Pi_k = \alpha\beta$, we specify $\beta$ matrix, say, $\beta^*$. 
\item[Step 3] Considering $\Pi_k$ time-varying, we estimate $\Delta\Pi_{k,t}$ for each period $t$.  
\item[Step 4] We obtain $\Pi_{k,t}$ through the cumulated sum of $\Delta\Pi_{k,t}$'s.
\item[Step 5] Supposing $\Pi_{k,t}=\alpha_t\beta^*$, we calculate $\alpha_t, t=1,\cdots$
\item[Step 6] We obtain our degree of comovement from $\alpha_t$'s in the same way as Ito et al.(2021).
\end{description}

We provide here practical information about our estimation. 
The sample size = 306; the number of variables = 6. 
According to \cite{ito2021time}, we choose $k=2$ in (\ref{eq: VECM basics}), the total number of parameters is 11016 when we attempt to obtain time-varying loading matrices to calculate our time-varying degree.   
We need to try many values of the hyperparameter of the group LASSO model to find some breaks. It takes about 1800 seconds on my Mac mini (Late 2014) with Intel Core i7 (16 GB 1600 MHz DDR3 memory).

Figures 3 and 4 summarize this paper's findings as Figures 1 and 2. Moreover, the results of breakpoint are summarized as follows.
\begin{figure}
 \centering
   \includegraphics[width=100mm]{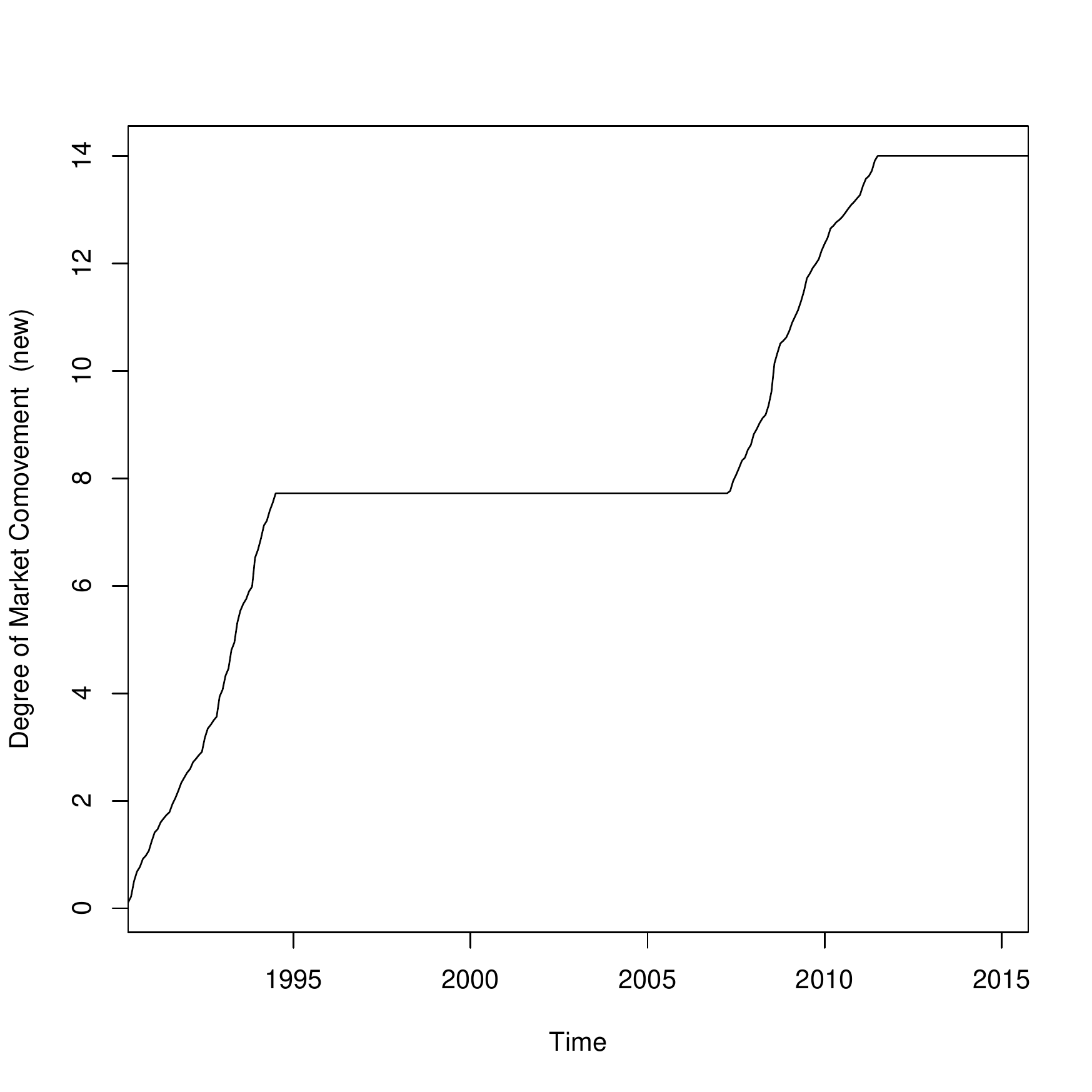}
 \caption{Degree of comovement in foreign exhange rates (1990-2015) }
 \label{fig:premium(new)}
\end{figure}

\begin{figure}
 \centering
   \includegraphics[width=100mm]{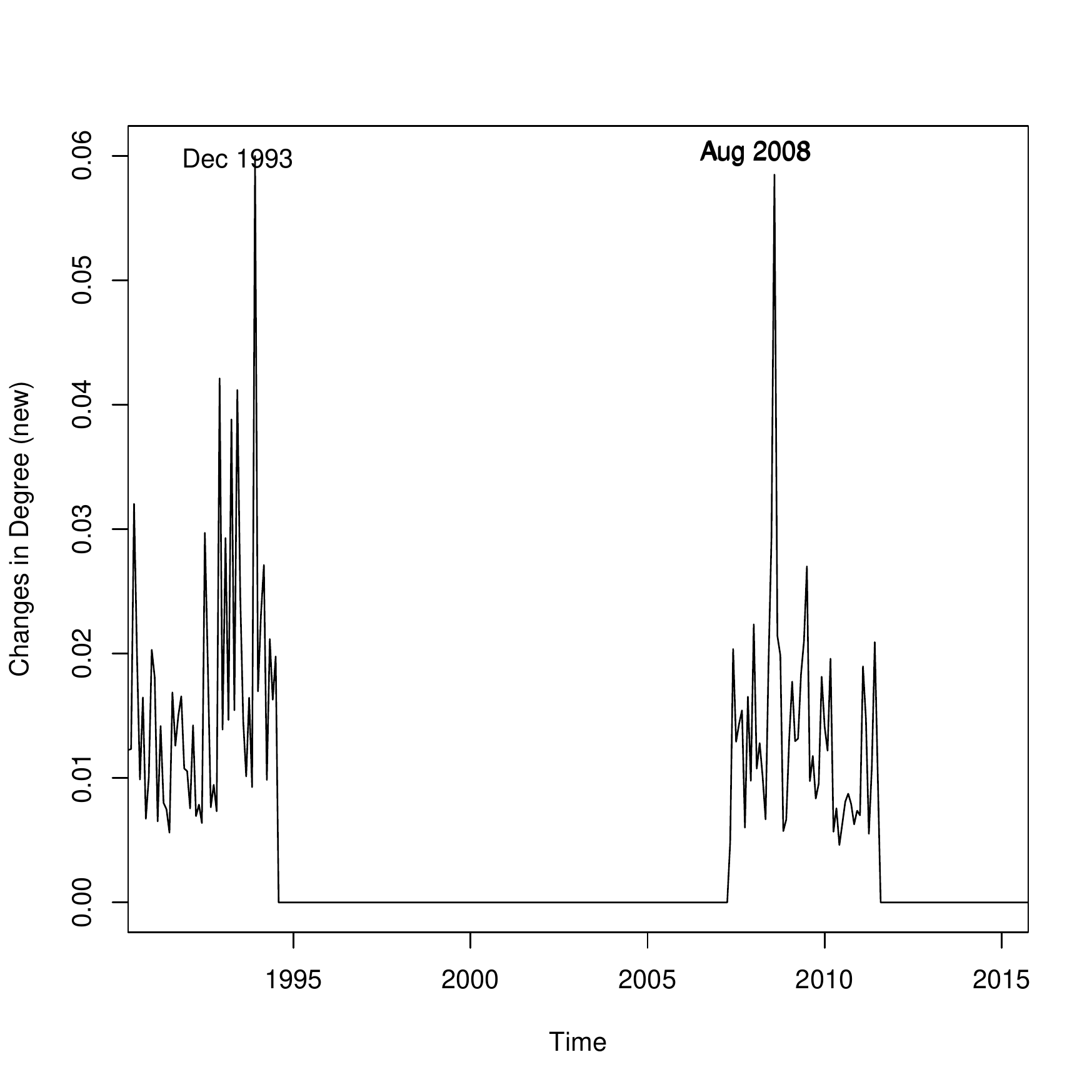}
 \caption{Changes in the Degree of comovement (1990-2015)}
 \label{fig:spot returns(new)}
\end{figure}

\begin{description}
\item[1st Impact] The degree is non-zero between May 1990 and July 1994 and attains a local maximum at December 1993.
\item[2nd Impact] The degree is non-zero between May 2007 and July 2011 and attains a local maximum at August 2008.
\end{description}

Notice that the first break possibly occurred before May 1990, and the left peak in the figure \ref{fig:spot returns(new)} possibly reflects the UK pound crisis in September 1992 as well as the UK's rapid economic recovery after the crisis. The right peak in the figure demonstrates Lehman Brothers Holdings Inc.'s bankruptcy on September 15, 2008.

\section{Concluding Remarks}
Here the author provides some remarks.  
We have shown a method to estimate linear time series models with time-varying parameters (TVP) that jump in very few periods. Using group LASSO allows us to detect breakpoints when addressing multivariate linear models.
The robustness relies on not any statistical technique but a simple, sparse technique that guarantees sparse structure parameters. 
This paper's result does not reflect the Asian financial crisis from July 1997. It could come from our choice of data: Canada, Japan, UK.
Ito et al.(2021) also detect two peaks of the changes in the Degree of comovement while the "resolution" is lower than this paper. The two estimates of Ito et al.(2021) and this paper, which estimate is better? It depends on what information the users want.

Finally, we show some future tasks: more statistical inference more applications to evaluate our framework's effectiveness. We should study correspondences between our detected breaks and possible actual shocks in practice. We are searching for better forecast performance by focusing on the periods after the {\it last} breakpoint by our technique.   

\newpage

\bibliography{LASSO2021} 
\bibliographystyle{asa}

\newpage

\section*{Appendix}
One can regard $\ve{\beta}_0$ as a prior if they are a Bayesian. But, I propose the least square estimate of the following linear regression model to obtain $\ve{\beta}_0$ conveniently with our givens.  

\begin{equation}   
\left[
\begin{array}{c}
    \ve{y}_{ 1 } \\
    \ve{y}_{ 2 } \\
    \ve{y}_{ 3 } \\
    \vdots   \\
    \ve{y}_{ T } 
\end{array}
\right]
= 
\left[
\begin{array}{c}
    X_{ 1 } \\
    X_{ 2 } \\
    X_{ 3 } \\
    \vdots   \\
    X_{ T } 
\end{array}
\right]
\ve{\beta}  
+
\left[
\begin{array}{c}
    \ve{\varepsilon}_{ 1 } \\
    \ve{\varepsilon}_{ 2 } \\
    \ve{\varepsilon}_{ 3 } \\
    \vdots  \\
    \ve{\varepsilon}_{ T } 
\end{array}
\right].
\label{eq:prior b}
\end{equation}



\end{document}